\documentclass[%
 reprint,
preprintnumbers,
 amsmath,amssymb,
 aps,
prb,
]{revtex4-2}

\usepackage{tabularx}
\usepackage{graphicx}
\usepackage{dcolumn}
\usepackage{bm}
\usepackage[font={small}]{caption}
\usepackage{subcaption}
\usepackage{color}
\usepackage{multirow}
\usepackage{xcolor}
\usepackage[export]{adjustbox}
\usepackage{url}

\newcommand{\fref}[1]{Figure~\ref{#1}}
\newcommand{\tref}[1]{Table~\ref{#1}}

\begin{document}

\renewcommand{\arraystretch}{1.2}
\preprint{\large SAND2022-3561 O }

\title{Electronic structure of $\boldsymbol{\alpha}$-RuCl$_3$ by fixed-node \\
and fixed-phase diffusion Monte Carlo methods}

\author{Abdulgani Annaberdiyev$^{1,3}$, Cody A. Melton$^{2}$, Guangming Wang$^{1}$, and Lubos Mitas$^1$}
\affiliation{
1) Department of Physics, North Carolina State University, Raleigh, North Carolina 27695-8202, USA \\}
\affiliation{
2) Sandia National Laboratories, Albuquerque, New Mexico 87123, USA \\}
\affiliation{
3) Center for Nanophase Materials Sciences Division, Oak Ridge National Laboratory, Oak Ridge, Tennessee 37831, USA \\}


\begin{abstract}

Layered material $\alpha$-RuCl$_3$ has caught wide attention due to its possible realization of Kitaev's spin liquid and
its electronic structure that involves the interplay of electron-electron correlations and spin-orbit effects. Several
DFT$+U$ studies have suggested that both electron-electron correlations and spin-orbit effects are crucial for accurately describing the band gap. 
This work studies the importance of these two effects using fixed-node and fixed-phase  diffusion Monte Carlo
calculations both in spin-averaged and explicit spin-orbit formalisms.
In the latter, the Slater-Jastrow trial function
is constructed from two-component spin-orbitals using 
our recent quantum Monte Carlo (QMC) developments and thoroughly tested effective core potentials.
Our results show that the gap in the ideal crystal is already accurately described by the spin-averaged case, with the dominant role being played by the magnetic ground state with significant exchange and electron correlation effects. We
find qualitative agreement between hybrid DFT, DFT+$U$, and QMC.
In addition, QMC results agree very well with available experiments, and we identify the values of exact Fock exchange
mixing that provide comparable gaps. Explicit spin-orbit QMC calculations reveal that the effect of spin-orbit coupling on the gap is minor, of the order of 0.2 eV, which corresponds to the strength of the spin-orbit of the Ru atom.
\end{abstract}

\maketitle


\section{Introduction}
\label{sec:intro}
Kitaev's seminal work \cite{kitaevAnyonsExactlySolved2006} on quantum spin liquids has sparked numerous experimental and
computational endeavors in search of candidate materials to realize this exotic phase
\cite{
plumbEnsuremathAlphaEnsuremath2014,
sandilandsSpinorbitExcitationsElectronic2016,
kimKitaevMagnetismHoneycomb2015,
sinnElectronicStructureKitaev2016,
caoLowtemperatureCrystalMagnetic2016,
johnsonMonoclinicCrystalStructure2015,
kimCrystalStructureMagnetism2016,
koitzschMathrmeffDescriptionHoneycomb2016,
reschkeElectronicPhononExcitations2017,
sandilandsOpticalProbeHeisenbergKitaev2016,
sarikurtElectronicMagneticProperties2018,
tianOpticallyDrivenMagnetic2019,
vatanseverStrainEffectsElectronic2019,
yadavKitaevExchangeFieldinduced2016,
zhouAngleresolvedPhotoemissionStudy2016,
ziatdinovAtomicscaleObservationStructural2016,
zhangTheoreticalStudyCrystal2021,
majumder_anisotropic_2015,
zhangTheoreticalStudyCrystal2021}.
In particular, despite the Ru atom's weak
spin-orbit coupling (SOC), $\alpha$-RuCl$_3$ has been suggested as a prime candidate. However, to properly analyze and describe the spin liquid physics with the meV energy scale (and potentially derive appropriate effective Hamiltonians), it is important to understand the basic electronic structure of this promising material. To this end,  numerous computational studies of
 $\alpha$-RuCl$_3$ have been carried out, mainly using Density Functional Theory (DFT)
with the Hubbard $U$ parameter (DFT$+U$) method \cite{plumbEnsuremathAlphaEnsuremath2014, zhangTheoreticalStudyCrystal2021, sandilandsSpinorbitExcitationsElectronic2016, kimKitaevMagnetismHoneycomb2015, sinnElectronicStructureKitaev2016}.  Interestingly, these studies resulted in data that did not provide an unambiguous picture, with lingering questions arising, especially on the origin of the band gap.  These DFT calculations essentially agreed on
the observation that it is a Mott insulator; however, the value of the gap and how it is impacted by SOC showed
significant differences between the studies. Obtained band gap estimates covered a significant range with values such as 
0.2~eV \cite{plumbEnsuremathAlphaEnsuremath2014},
0.4~eV \cite{majumder_anisotropic_2015},
0.7~eV \cite{zhangTheoreticalStudyCrystal2021},
1.0~eV \cite{sandilandsSpinorbitExcitationsElectronic2016, kimKitaevMagnetismHoneycomb2015},
1.2~eV \cite{koitzschMathrmeffDescriptionHoneycomb2016},
and 1.9~eV \cite{sinnElectronicStructureKitaev2016}.
In particular, some studies indicated that it is a spin-orbit assisted Mott
insulator with non-negligible SOC effects \cite{plumbEnsuremathAlphaEnsuremath2014, sinnElectronicStructureKitaev2016}. The matter is further complicated by the occurrence of  optically generated excitons with sizable binding energies that make the analysis of the fundamental band gap much less straightforward.

In this work, we aim to address these points using
many-body fixed-node and fixed-phase diffusion Monte Carlo
(DMC) methods. 
Specifically, we carry out a pioneering study of the band gap in the solid with the explicit contribution of SOC using
both spin-orbit averaged fixed-node DMC (FNDMC) and two-component-spinor fixed-phase spin-orbit
DMC (FPSODMC) calculations
\cite{kolorencApplicationsQuantumMonte2011,meltonManybodyElectronicStructure2020}.  
We carefully verify the valence-only Hamiltonians represented by
effective core (pseudo)potentials (ECPs) with a proper account of spin-orbit effects.
Systematic errors that could significantly affect the quality of results, such as finite-size effects, basis sets, and fixed-node biases are analyzed. We focus on studying a periodic,
stoichiometric, ideal model without considering defects, impurities, or localized excitonic effects.  Our many-body trial functions are single-reference that proved successful in previous studies of transition metal oxides with significant electron-electron correlations \cite{kolorencApplicationsQuantumMonte2011}.

A very good agreement between fundamental and promotion gaps is obtained with a value close to $\approx $ 2 eV, which agrees very well with photoemission experiments. Most importantly,
we find that the basic electronic structure
parameters such as cohesion and band gap are mainly determined by spin-averaged physics. 
The impact of spin-orbit on these characteristics is found to be rather minor, confined to $\approx$ 0.1--0.2~eV (5--10\%) shifts in the band gap and $\approx$ 0.5--0.6~eV (4--5\%) shifts in cohesion.
Therefore, we show that the key issue of $\alpha$-RuCl$_3$ is the electron correlation that should be properly included in any realistic study of low-lying states of putative quantum spin liquids.
To the best of our knowledge, we present the first two-component spinor many-body wave function calculation
of a solid that recovers more than 95\% of the valence correlation energy in a variational setting.


The paper is organized as follows. We begin with RuCl$_3$ crystal structure (Sec. \ref{sec:structure}) and methods sections (Sec. \ref{sec:methods}) that include brief
descriptions of employed DFT and QMC approaches. The results (Sec. \ref{sec:results}) involve testing the accuracy of pseudopotentials, probing
the finite-size and fixed-node errors, calculating the cohesive energies and gaps with the explicit impact of the spin-orbit
effects. We conclude (Sec. \ref{sec:conclusions}) by pointing out the salient results and future perspectives of the presented methodological
improvements.

\section{Crystal Structure}
\label{sec:structure}
$\alpha$-RuCl$_3$ consists of weakly coupled 2D layers of RuCl$_6$ octahedra.
The Ru atoms form a near-perfect hexagonal structure and sit in an almost-ideal, edge-sharing RuCl$_6$ octahedron.

Throughout this work, we used the experimentally measured geometry from Ref.~\cite{banerjeeProximateKitaevQuantum2016}.
There are two proposed space-group symmetries for this system \cite{banerjeeProximateKitaevQuantum2016}: $C2/m$ (No. 12)
and $P3_112$ (No. 151), which differ mainly in their stacking order. We used the $C2/m$ structure throughout the work since
it is suggested to be the correct ground state from experimental observations
\cite{caoLowtemperatureCrystalMagnetic2016, johnsonMonoclinicCrystalStructure2015} and from DFT$+U$ calculations \cite{kimCrystalStructureMagnetism2016}. We
also note that the energy differences between the above space-group symmetries are rather small ($\sim$1~meV/Ru) as
found previously \cite{kimCrystalStructureMagnetism2016}.

\begin{figure}[!htbp]
\centering
\begin{subfigure}{0.5\columnwidth}
\includegraphics[width=\columnwidth]{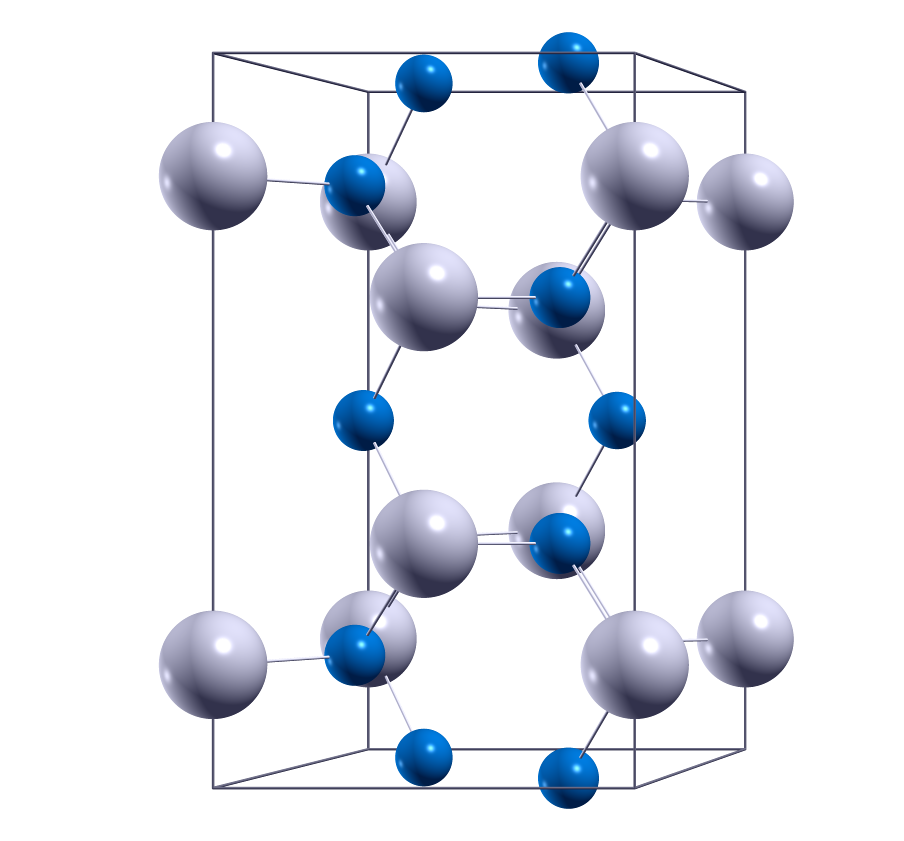}
\caption{
XY plane
}
\end{subfigure}%
\begin{subfigure}{0.5\columnwidth}
\includegraphics[width=\columnwidth]{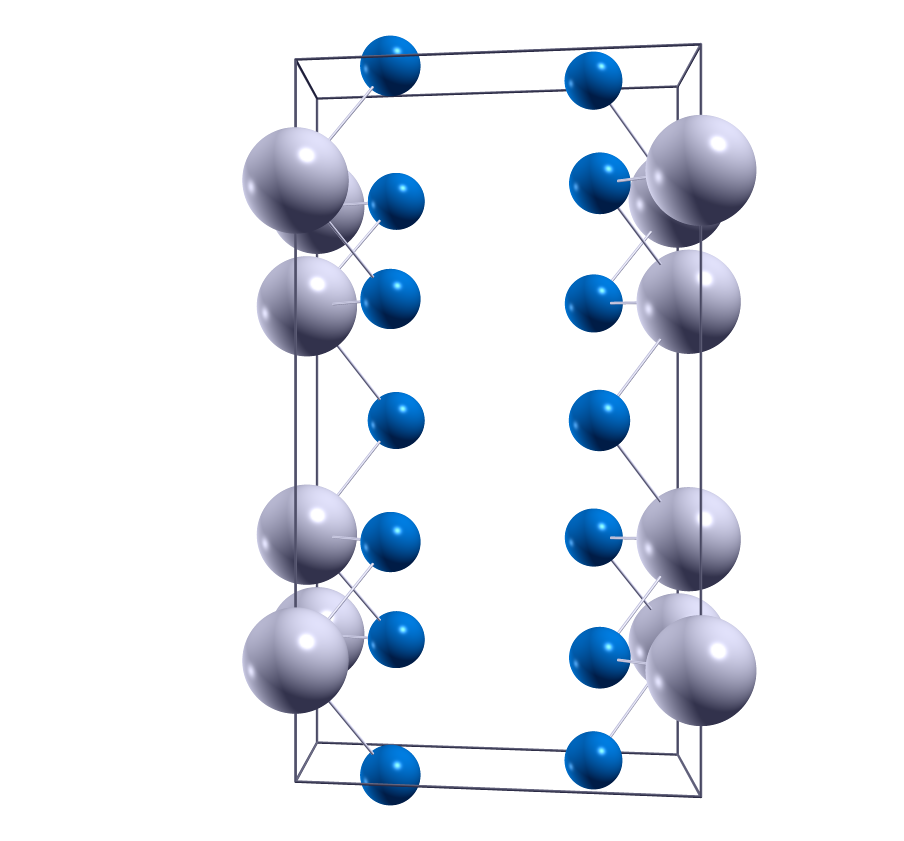}
\caption{
YZ plane
}
\end{subfigure}
\caption{
Conventional cell crystal structure of $\alpha$-RuCl$_3$.
(a) XY plane and (b) YZ plane views are shown.
Large spheres with light colors are Ru atoms, and small spheres with dark colors are Cl atoms.
\label{fig:111_cell}
}
\end{figure}

A primitive cell of $C2/m$ symmetry corresponds to a monoclinic cell with two Ru atoms.
Although this cell is large enough to describe the ferromagnetic (FM) phase, for the zigzag antiferromagnetic (ZZ) phase, it has to expand to the conventional cell with four Ru atoms since antiferromagnetic order breaks the equivalency of the Ru atoms.
Therefore, the conventional cell (Fig. \ref{fig:111_cell}) has been used for DFT calculations and to obtain the larger QMC supercells via periodic tiling. 

\section{Methods}
\label{sec:methods}

We use effective core potentials (ECPs) to eliminate the core electrons and to make the calculations feasible.
Specifically, we will refer to two types of ECPs in this work. One is the averaged spin-orbit relativistic effective
potential (AREP), which accounts for only scalar relativity and therefore does not include explicit spin-orbit coupling
terms. The other ECP is a spin-orbit relativistic effective potential (SOREP), which is fully relativistic with explicit
two-component spin-orbit terms. The AREP/SOREP labels will indicate whether explicit spin-orbit coupling is included or
not in the following sections.

The orbitals/spinors are obtained from Kohn-Sham DFT using the \textsc{Quantum ESPRESSO} code
\cite{giannozziQUANTUMESPRESSOModular2009}. We used DFT$+U$ with atomic wave function projectors and hybrid-DFT with exact Fock exchange formulations where
PBE is used as the underlying functional in both cases. In this work, $\omega$ represents the exact Fock exchange weight
in hybrid DFT functional with exchange and correlation components:
\begin{equation}
\label{eqn:hybrid_def}
{\rm PBE0}(w) = w {\rm E^{HF}_x} + (1-w) {\rm E^{PBE}_x} + {\rm E^{PBE}_c}
\end{equation}
where $\omega$ will be expressed in percentages throughout the paper. 
The weight $w=25\%$ corresponds to the definition of $\mathrm{PBE0}$ functional \cite{perdewRationaleMixingExact1996, adamoReliableDensityFunctional1999}.
In all SCF calculations, we used $400$~Ry kinetic energy
cutoff and $2\times2\times2$ $k-$mesh in the conventional cell. We have tested that these parameters result in converged
energies and gaps, which can be found in the Supplemental Material.

We use single-reference Slater-Jastrow trial wave functions throughout the work except for a few special calculations
for the Ru atom, where we used the complete open-shell configuration interaction (COSCI) method to build the appropriate
symmetry multi-reference states with spinors. For clarity, the calculations are labeled as  
(QMC method)/(trial wave function orbitals), such as FNDMC/PBE, etc. QMC calculations are carried out using
\textsc{QMCPACK}  \cite{kimQMCPACKOpenSourceab2018, kentQMCPACKAdvancesDevelopment2020} code. Unless otherwise specified, all
DMC calculations with hybrid DFT orbitals are timestep extrapolated using [0.02, 0.005] Ha$^{-1}$ timesteps (with
a timestep ratio of 4 as suggested in Ref. \cite{leeStrategiesImprovingEfficiency2011}) and use the T-moves algorithm
\cite{casulaSizeconsistentVariationalApproaches2010} for pseudopotential evaluation. Calculations using orbitals from
DFT$+U$ calculations use a 0.02 Ha$^{-1}$ timestep and the locality approximation
\cite{mitasNonlocalPseudopotentialsDiffusion1991} for pseudopotential evaluation. In cases where there is a direct
comparison of DMC/(hybrid DFT) vs. DMC/DFT$+U$ results, the same 0.02 Ha$^{-1}$ timestep and T-moves are used
consistently.

In the standard FNDMC approach, the ground state wave function and energies are obtained through an imaginary time
evolution process from an initial trial wave function. Under specific boundary conditions (open or special $k-$points in
periodic boundary conditions), the wave function is real-valued, and the fermion sign problem is controlled through the
fixed node approximation where the final nodes must match the trial wave function's nodal structure. Additionally, since
the Hamiltonian does not depend on the spin in the absence of spin-orbit, the electron spin is conserved, and each
electron spin is fixed throughout the simulation. In the FPSODMC, where the Hamiltonian now includes spin-orbit
coupling, the spin degree of freedom needs to be sampled since spin is no longer conserved. 
To facilitate sampling, a smooth and complex representation for the spin is introduced, which makes the wave function complex.
In this case, the fixed-phase approximation is introduced where the final phase of the wave function must match the trial wave
function. See the original papers on this topic for a thorough discussion \cite{melton_quantum_2016,
melton_fixed-node_2016, melton_spin-orbit_2016, melton_quantum_2017, melton_projector_2019}. 

\section{Results}
\label{sec:results}

\subsection{Accuracy of Pseudopotential Hamiltonians}
\label{sub:ecp}

Let us elaborate on the accuracy of employed ECPs since this will enable us to validate the accuracy of calculations and
check for possible biases. For the Cl atom, we use an AREP type, correlation consistent effective core potential (ccECP)
with 7 valence electrons \cite{bennettNewGenerationEffective2018}. The high accuracy of this ECP was demonstrated in the
original paper,
where the errors in the low-lying atomic gaps, such as ionization potential and electron affinity, were
smaller than chemical accuracy (1 kcal/mol $\approx 0.043$ eV). Additionally, the binding energies of ClO and Cl$_2$ molecules
across various bond geometries agreed with all-electron (AE) CCSD(T) within the chemical accuracy boundary. 
We neglect the spin-orbit coupling effects for this element, but we account for scalar relativity, which is implicitly included in ccECP. 
The interested reader is referred to the original paper for more data and thorough discussions of the Cl ECP
\cite{bennettNewGenerationEffective2018}.

\begin{table}[!htbp]
\centering
\caption{Ru atom scalar relativistic AE excitations and corresponding errors for various core approximations. ECPs
correspond to AREP with 16 valence electrons. All calculations are performed with the RCCSD(T) method using an
uncontracted aug-cc-pCV5Z basis set. All values are in eV.}
\label{tab:ecp_arep_gaps}
\begin{tabular}{llr|ccccc}
\hline
States                  & Term         &  AE     &  MDFSTU & reg-MDFSTU &   UC \\
\hline
{[Kr] $4d^{7}5s^{1}$ }  & $^5$F        &   0.000 &         &          &        \\
{[Kr] $4d^{7}5s^{2}$ }  & $^4$F        &  -1.022 &   0.007 &    0.007 &  0.007 \\
{[Kr] $4d^{9}$       }  & $^2$D        &   2.383 &   0.033 &    0.034 & -0.051 \\
{[Kr] $4d^{8}$       }  & $^3$F        &   1.215 &   0.034 &    0.033 & -0.034 \\
{[Kr] $4d^{6}5s^{2}$ }  & $^5$D        &   0.810 &  -0.045 &   -0.045 &  0.033 \\
{[Kr] $4d^{7}$       }  & $^4$F        &   7.396 &  -0.010 &   -0.010 & -0.023 \\
{[Kr] $4d^{6}5s^{1}$ }  & $^6$D        &   8.377 &  -0.059 &   -0.059 &  0.021 \\
{[Kr] $4d^{6}$       }  & $^5$D        &  24.003 &  -0.068 &   -0.068 & -0.007 \\
{[Kr] $4d^{5}$       }  & $^2$T$_{2g}$ &  57.044 &  -0.002 &   -0.002 & -0.007 \\
\hline
MAD                     &              &         &   0.032 &    0.032 &  0.023 \\


\hline
\end{tabular}
\end{table}

For the Ru atom, we use the small-core ECP of the Stuttgart group (MDFSTU) with 16 valence electrons, and it is given
both in AREP and SOREP forms \cite{petersonEnergyconsistentRelativisticPseudopotentials2007}. Stuttgart group ECPs have a
$-Z_{\rm eff}/r$ Coulomb singularity at the origin, making it problematic  to use in DFT plane-wave calculations due to
resulting high kinetic energy cut-offs and more costly in QMC calculations due to an increase in local energy fluctuations.
To overcome these issues, we have modified the MDFSTU ECP by adding a regularizing local potential that cancels
out the Coulomb singularity near the ionic origin (reg-MDFSTU, see Supplemental Material for ECP parametrization). This
has been done in a very conservative manner, i.e., only at the vicinity of $r=0$ since that preserves properties of the
original construction. \tref{tab:ecp_arep_gaps} demonstrates the accuracy of the AREP part of the reg-MDFSTU and shows
that the spectrum is essentially the same as the original MDFSTU ECP. The spectrum is calculated using the RCCSD(T)
method for AE and ECP cases and the mean absolute deviation (MAD) is smaller than the chemical accuracy.
Here UC stands for an uncorrelated core, where the underlying self-consistent calculation is fully relaxed, and the
subsequent correlated calculations do not allow for any excitations that involve those core states. We can see that
reg-MDFSTU is only slightly worse than UC, therefore, indicating the high quality of this ECP. In
\tref{tab:ecp_arep_gaps}, $^2$T$_{2g}$ state represents the triply degenerate $t_{2g}$ states encountered in solid state
calculations with octahedral geometry. This state occupies $d_{xy}$, $d_{yz}$, $d_{xz}$ orbitals, with one of these left
singly occupied and others doubly occupied. The accuracy of this state is relevant to the solid calculations where the
Ru atoms exhibit the nominal Ru$^{+3}$ ($d^5$) occupation in $t_{2g}$ orbital. Note that this ``bare" atomic excitation
is nominally very high compared to what is expected in a crystalline environment since hybridization and crystal field
pulls it significantly lower. Despite its atomic value being over 50 eV, it is remarkable that it is reproduced with the
same accuracy as the other atomic low-lying states. From the data in \tref{tab:ecp_arep_gaps}, we can conclude that the
AREP accuracy of the employed ECP is sufficiently high for further correlated QMC calculations.

In \tref{tab:j_splits}, we show the accuracy of spin-orbit terms of SOREP using explicit spin-orbit calculations. In
this case, we directly compare with experiments to better elucidate the expected errors for the employed ECPs. Here the
spinors are obtained from the \textsc{Dirac} code \cite{saueDIRACCodeRelativistic2020}, and COSCI is used to obtain the
multi-reference states. We can see that both the COSCI and QMC methods using COSCI as trial wave functions result
in minor errors that are well within the other systematic errors present in this work. We have thus established that
both AREP and SOREP type approximations are accurate, allowing us to study and estimate other systematic biases present
in our calculations.

\begin{table}[!htbp]
\centering
\caption{
Ru atom experimental J-splitting excitations and corresponding errors using various methods.
All values are in eV.
}
\label{tab:j_splits}
\begin{tabular}{cc|ccccc}
\hline
Term    &  Expt. \cite{NIST_ASD} &         COSCI  & VMC/COSCI   & DMC/COSCI   \\
\hline
$^5F_5$ & 0.0000 &         0.0000 &     0.0000  &     0.0000  \\
$^5F_4$ & 0.1476 &         0.0349 &     0.02(2) &     0.08(2) \\
$^5F_3$ & 0.2593 &         0.0597 &     0.07(2) &     0.07(2) \\
$^5F_2$ & 0.3364 &         0.0742 &     0.05(2) &     0.05(2) \\
$^5F_1$ & 0.3850 &         0.0823 &     0.04(2) &     0.07(2) \\
\hline
MAD     &        &         0.0628 &     0.05(1) &     0.07(1) \\
\hline
\end{tabular}
\end{table}

\subsection{DFT Functionals, Orbitals and Band Structures}
\label{sub:bands}

Previous DFT$+U$ calculations found that $\alpha$-RuCl$_3$ features almost degenerate magnetic states
\cite{kimKitaevMagnetismHoneycomb2015}. 
The ferromagnetic phase (FM) and the anti-ferromagnetic zigzag phase (ZZ) are two main candidate ground state phases.
Most calculations and measurements \cite{banerjeeProximateKitaevQuantum2016, johnsonMonoclinicCrystalStructure2015} suggest that the ZZ phase is the genuine ground state, although DFT$+U$ predicts the FM phase to be lower in energy for some values of $U$
\cite{kimKitaevMagnetismHoneycomb2015}. 
To verify this system's correct ground state, we performed FNDMC calculations using trial wave functions based on hybrid-DFT PBE0 functional as shown in Table \ref{tab:compare_states}.
Since the ZZ phase results in the lowest energy, we focus on this magnetic state in what follows. 
Note that at the
DFT$+U$ level, the energy difference is very small between the FM and ZZ phases ($\approx 1$ meV/Ru)
\cite{kimKitaevMagnetismHoneycomb2015}, whereas DMC predicts a much larger difference of $\approx 0.07(1)$ eV/Ru.

\begin{table}[!htbp]
\centering
\caption{Total FNDMC energies with trial functions based on orbitals from PBE0 with 15\% exact exchange and averaged
spin-orbit AREP formalism. Conventional cell energies [Ha] in nonmagnetic (NM), ferromagnetic (FM), and zig-zag
antiferromagnetic (ZZ) phases are shown. Note significant gain from magnetic ordering.}
\label{tab:compare_states}
\begin{tabular}{c|c}
\hline
Phase    & FNDMC Energy  \\
\hline
NM & -558.012(1)  \\
FM & -558.900(1)  \\
ZZ & -558.909(1) \\
\hline
\end{tabular}
\end{table}

\begin{figure*}[!htbp]
\centering
\begin{subfigure}{0.45\textwidth}
\includegraphics[width=\textwidth]{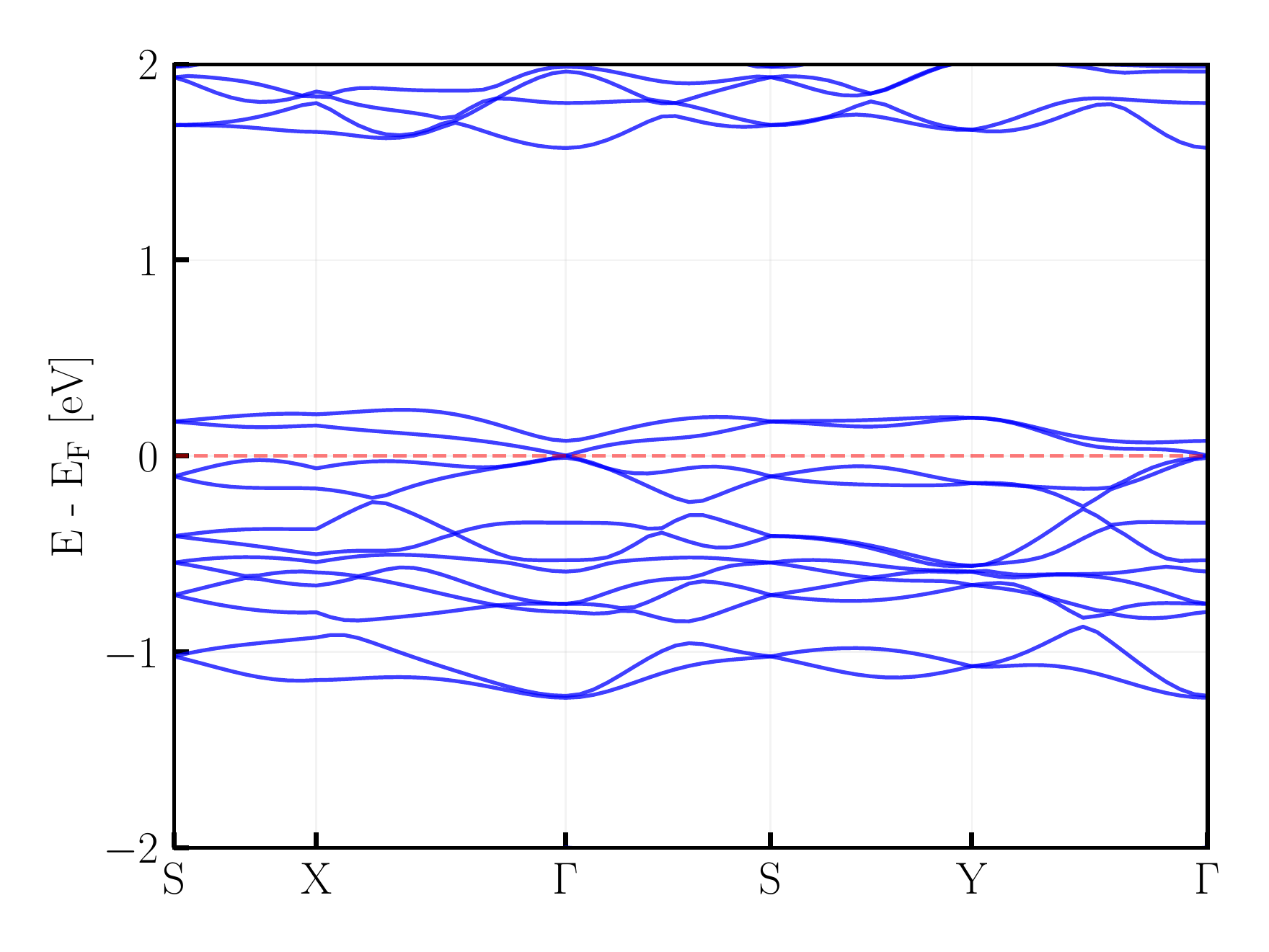}
\caption{
}
\end{subfigure}%
\begin{subfigure}{0.45\textwidth}
\includegraphics[width=\textwidth]{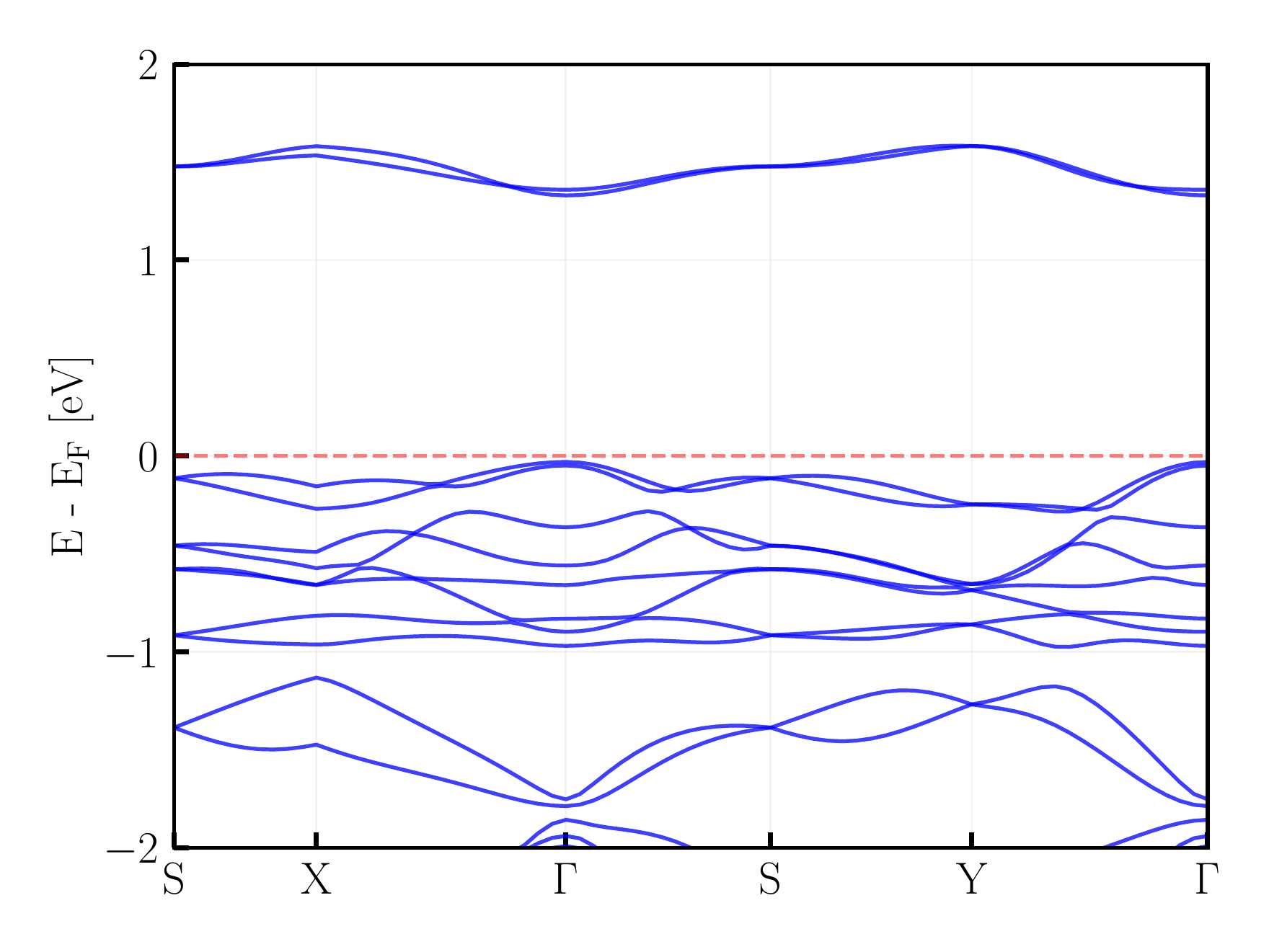}
\caption{
}
\end{subfigure}
\begin{subfigure}{0.45\textwidth}
\includegraphics[width=\textwidth]{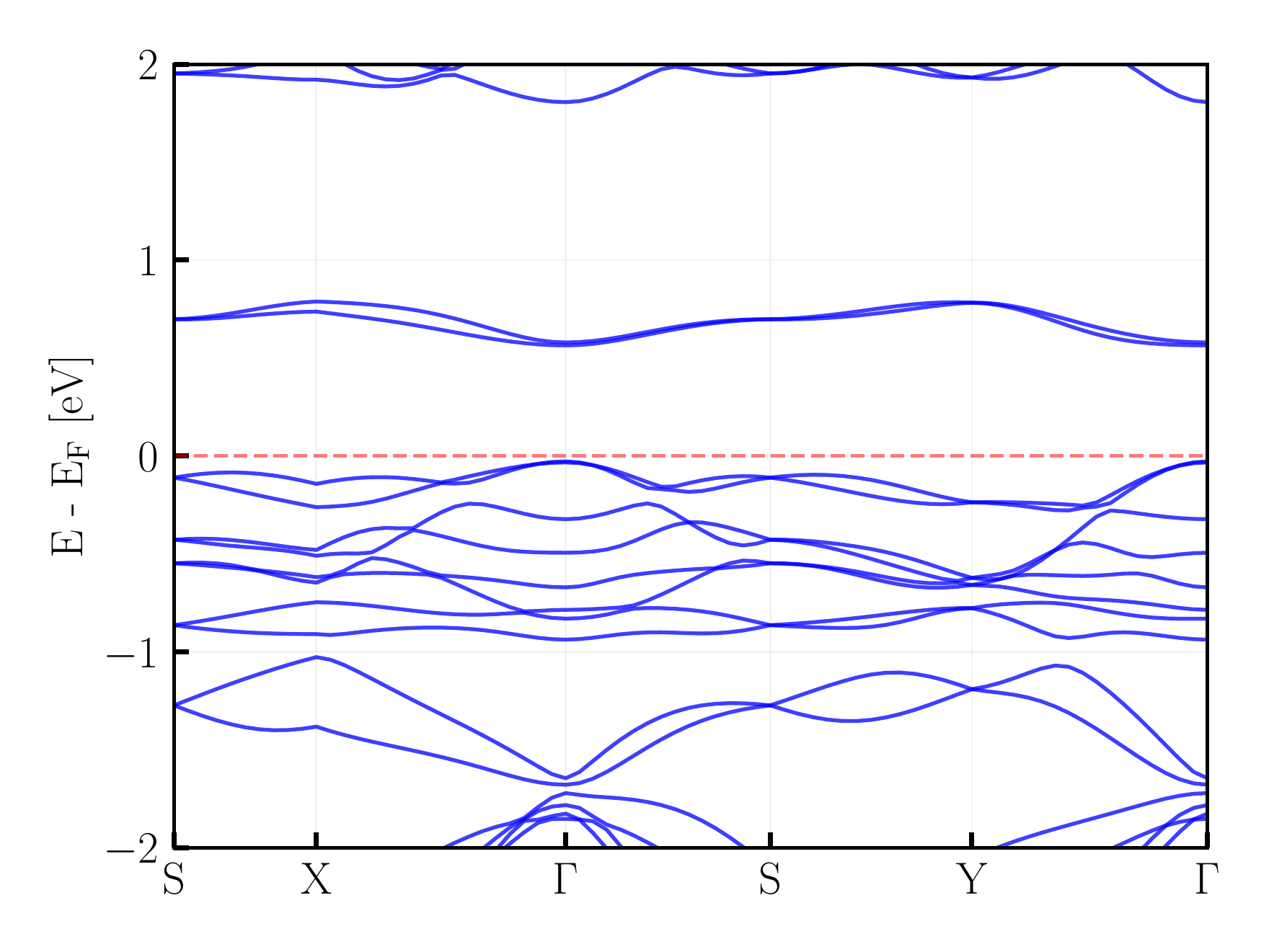}
\caption{
}
\end{subfigure}%
\begin{subfigure}{0.45\textwidth}
\includegraphics[width=\textwidth]{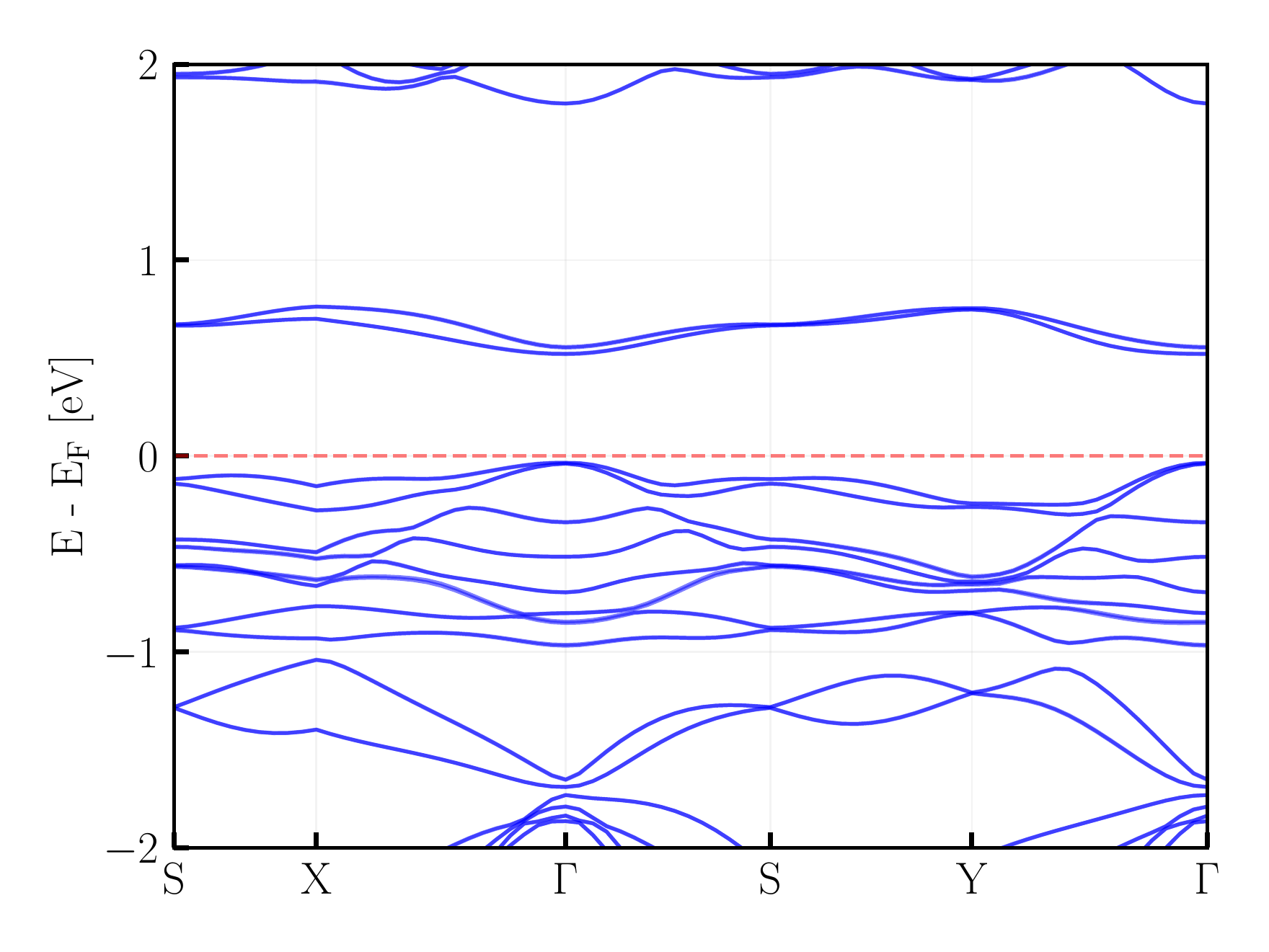}
\caption{
}
\end{subfigure}%
\caption{
Band structures of the conventional cell in the ZZ phase.
Employed functionals using spin-averaged AREP are plotted as follows:
(a) PBE, 
(b) PBE0(15\%),
(c) PBE$+U$(1.5~eV). Bands in the full SOREP framework  with PBE$+U$(1.5~eV)
are plotted in (d).
}
\label{fig:arep_bands_compare}
\end{figure*}

In \fref{fig:arep_bands_compare}, we plot the band structure corresponding to the system's conventional cell.
The plots show the band structures for PBE, PBE$+U$(1.5 eV), and PBE0(15\%) functionals, where the $+U$ and $\omega$
parameters are chosen based on FNDMC calculations. The choice for these specific values will be discussed later. Using
plain PBE, we observe the usual band gap underestimation with the resulting metallic state as it was observed before
\cite{plumbEnsuremathAlphaEnsuremath2014, sandilandsSpinorbitExcitationsElectronic2016, kimKitaevMagnetismHoneycomb2015,
sinnElectronicStructureKitaev2016}. 
However, as expected, reasonable values of $U$ and $\omega$ lead to finite gaps. 
Note that CBM to VBM transition corresponds to $\Gamma \to \Gamma$ transition; therefore, we study the $\Gamma \to \Gamma$ gap in the sections below.

\subsection{Estimation of Fixed-Node Bias}

\tref{tab:fn_bias_atoms} shows the estimation of FN bias for Ru and Cl atoms. For calculation of exact atomic energies,
we used CCSD(T) with extrapolation to the complete basis set (CBS) limit (for methodology, see
Ref. \cite{annaberdiyevAccurateAtomicCorrelation2020}). We can see that using a single-reference FNDMC calculation, we
can obtain accurate correlation energies with approximately $4\%$ and $3\%$ correlation energies missing for Ru and Cl, respectively. These calculations serve as an estimate for the expected FN bias in the solid calculations. We see that the Ru
atom's FN bias is much smaller than the isovalent Fe atom, which has $\eta = 10.5(1)\%$ error
\cite{annaberdiyevAccurateAtomicCorrelation2020}. The corresponding biases are significantly smaller since the $4d$
states are more delocalized than the corresponding $3d$ states, as pointed out previously
\cite{raschCommunicationFixednodeErrors2014}. These observations motivate us to move forward with
single-reference trial wave functions for solid calculations.

\begin{table}[!htbp]
\centering
\caption{Estimation of FN bias for Ru and Cl atoms. Total energies [Ha] using various methods are shown.
$\epsilon$~[mHa] represents the DMC/PBE0 error compared to CCSD(T)/CBS. $\eta$ represents the missing percentage of the
correlation energy $\eta = 100\% \cdot \epsilon / |E_{corr}|$.}
\label{tab:fn_bias_atoms}
\begin{tabular}{l|lll}
\hline
Qty.             &       Ru      &       Cl       \\
\hline
ROHF/CBS         &  -93.80594(5) &  -14.68947(1)  \\
CCSD(T)/CBS      &   -94.4892(9) &   -14.9267(2)  \\
FNDMC/PBE0         &   -94.4613(2) &   -14.9192(1)  \\
\hline
$\epsilon$ [mHa] &       27.9(9) &        7.5(2)  \\
$\eta$ [\%]      &        4.1(1) &       3.17(8)  \\

\hline
\end{tabular}
\end{table}

\subsection{Minimization of Fixed-Node Bias}
\label{sub:fn_bias}

One of the significant systematic errors present in QMC calculations is the fixed-node bias. Although direct
optimization of the trial wave functions to alleviate this is possible \cite{zhaoVariationalExcitationsReal2019,
townsendStartingpointindependentQuantumMonte2020, assaraf_optimizing_2017, filippi_simple_2016}, we take a more simple but rather effective approach of optimizing the
effective single-particle Hamiltonian, namely, the optimization of the hybrid DFT functional that results in the lowest
FNDMC energies.
This has the advantage of providing insight into one-particle effective theories and their indirect optimality test concerning subsequent QMC runs, and can be straightforwardly applied to complex solids.
As mentioned previously, we explore two different routes, hybrid functionals and the DFT$+U$ method. In the DFT$+U$ method,
the value of effective $U$ is varied while the exact Fock exchange factor $\omega$ is varied in hybrid-DFT with the goal
to find the best variational energy minimum.
We use the conventional cell at $k=\Gamma$ point for these nodal optimization calculations and probe the nodes of four
different states: ground state (GS), $\Gamma \to \Gamma$ promoted excited state, cation state (CA), and anion state
(AN). For simplicity, we will plot the average of these states, while the individual state data and plots can be found in
the Supplemental Material.
\begin{figure*}[!htbp]
\centering
\begin{subfigure}{0.5\textwidth}
\includegraphics[width=\textwidth]{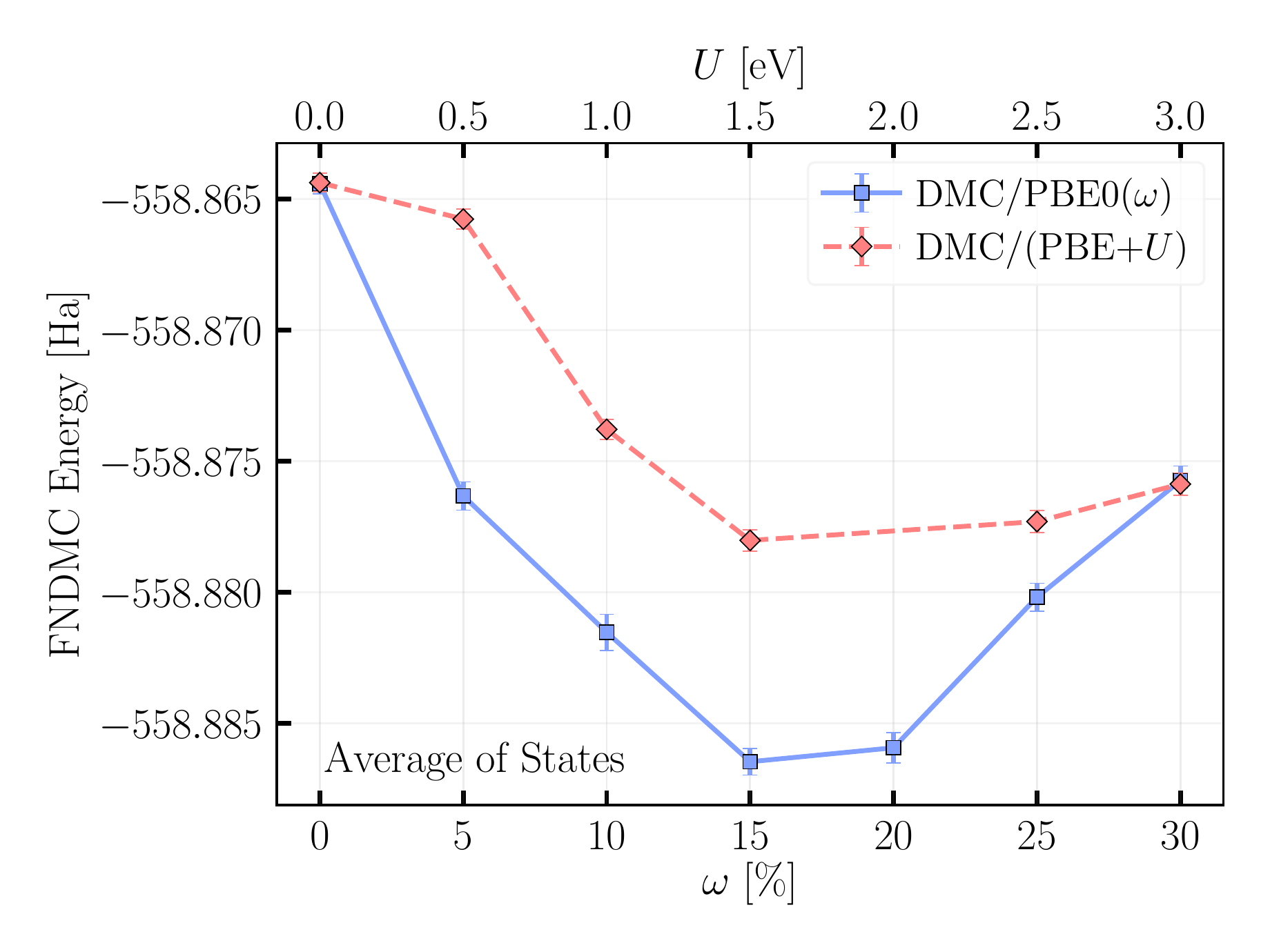}
\caption{
}
\label{fig:dmc_scan_avg_tot}
\end{subfigure}%
\begin{subfigure}{0.5\textwidth}
\includegraphics[width=\textwidth]{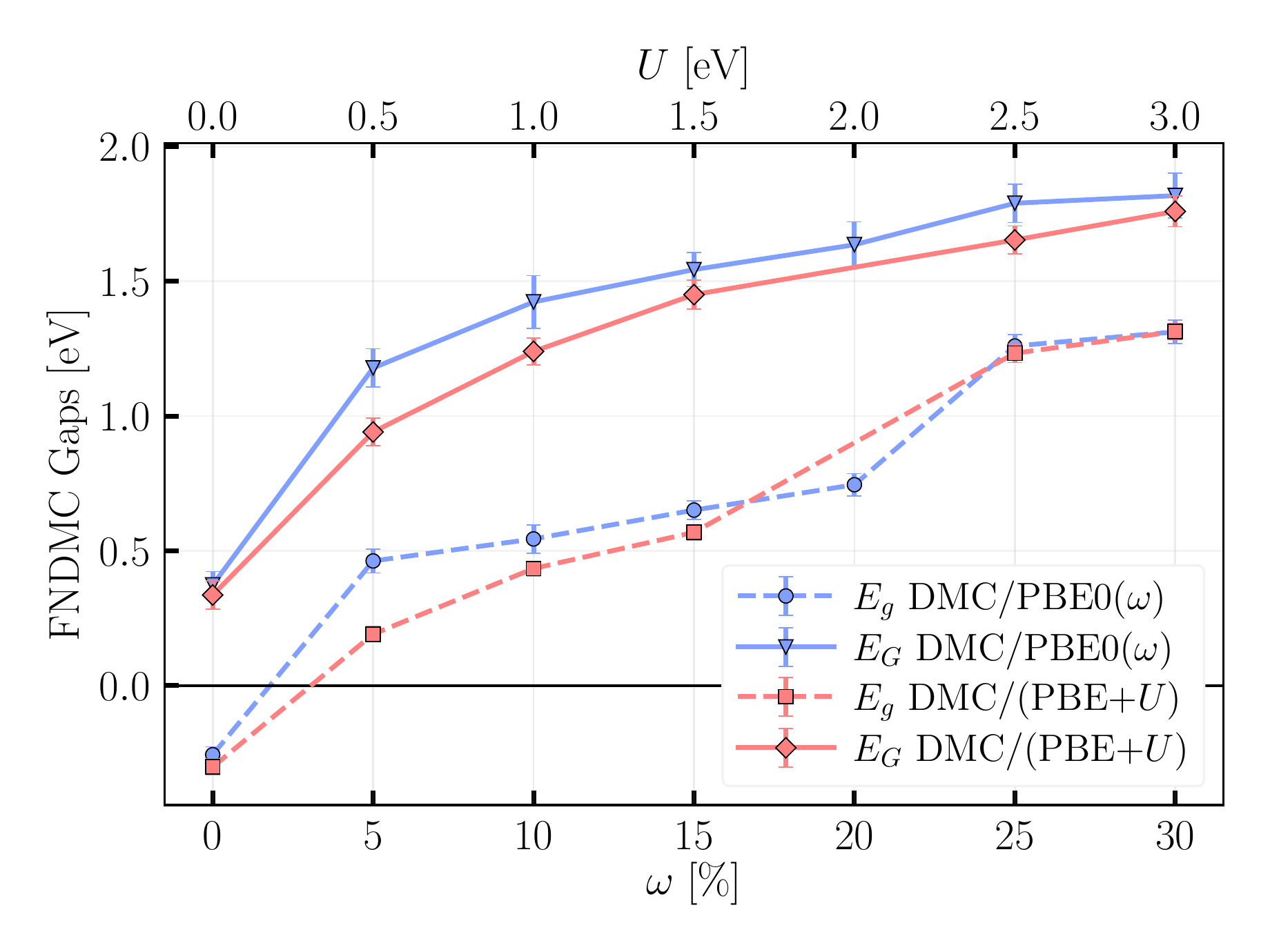}
\caption{
}
\label{fig:dmc_scan_avg_gap}
\end{subfigure}
\caption{FNDMC total energies and gaps for various nodal surfaces using exact exchange mixing $\omega$ and Hubbard $U$ parameters. 
FNDMC timestep \mbox{0.02 Ha$^{-1}$} and T-moves were used for these calculations. Energies correspond to
conventional cell ($n=4$ RuCl$_3$ units) in ZZ phase. (a) Average energies of all considered states. (b) Optical ($E_g$)
and fundamental ($E_G$) gaps.}
\end{figure*}
\fref{fig:dmc_scan_avg_tot} shows the average energies of the above-mentioned states. We can observe that there is a clear minimum in DMC/PBE0($\omega$) calculations around \mbox{$\approx 15\%$}. DMC/PBE$+U$ shows a minimum around
$U=1.5$~eV; however, the corresponding total energies are significantly higher. For the following calculations, we take
DMC/PBE0($15\%$) as the most optimal value for both ground and excitations, whereas the DMC/PBE$+U$(1.5 eV) will be used
as a secondary reference. Note that plots as \fref{fig:dmc_scan_avg_tot} comparing the nodes of hybrid-DFT and DFT$+U$
were presented also elsewhere with similar results \cite{bennettOriginMetalInsulatorTransitions2022}.

In \fref{fig:dmc_scan_avg_gap}, the calculated DMC gaps are shown for several trial wave functions and settings.
In particular, one can calculate the optical (or excitonic) gap as:
\begin{equation}
    \label{eqn:optical_gap}
    E_g = E^{\Gamma\Gamma}_{N} - E^{GS}_{N}
\end{equation}
and the fundamental (charge/quasi-particle) gap as:
\begin{equation}
    \label{eqn:fundamental_gap}
    E_G = IP - EA = E^{AN}_{N+1} + E^{CA}_{N-1} - 2 \cdot E^{GS}_{N}
\end{equation}
where $N$ is the number of electrons. \fref{fig:dmc_scan_avg_gap} shows the importance of obtaining proper trial wave
functions since both $E_g$ and $E_G$ heavily depend on the quality of the employed orbitals. Both types of trial wave
functions result in similar $E_g$, $E_G$ gaps, although DMC/PBE$+U$ are higher in total energy. Note that these values
are still tentative since they correspond to a rather small conventional cell with significant finite-size biases. It is
therefore important to filter out the presence of systematic shift in \fref{fig:dmc_scan_avg_gap} and to converge the
estimators with respect to the system size as in Section \ref{sub:finite_size}.

\subsection{Effective Charges and Magnetic Moments for QMC-Optimized $U/\omega$}

The VMC and DMC variationally optimized results also provide the best effective one-particle theory on post-DFT optimization space (Hubbard $U$ or Fock exchange weight).
Therefore it is worth probing for straightforward quantities such as effective charges and magnetic moments.
We provide the corresponding values in Tab \ref{tab:charges} and Tab \ref{tab:moments} respectively.   
\begin{table}[!htbp]
\centering
\caption{
Effective charges of Ru and Cl atoms in $\alpha$-RuCl$_3$ from DFT calculations (this work).
}
\label{tab:charges}
\begin{tabular}{l|cc}
\hline
Method & Ru ($e^-$) & Cl ($e^-$) \\
\hline
PBE             & 0.943 & -0.313 \\
PBE0(15\%)      & 1.022 & -0.342 \\
PBE0(20\%)      & 1.048 & -0.351 \\
PBE$+U$(1.5~eV) & 1.182 & -0.315 \\
\hline
Ref. \cite{setyawan_high-throughput_2010, aflow} 
                & 1.13 & -0.38 \\
\hline
\end{tabular}
\end{table}

Indeed, even these basic expectation values vary significantly between various flavors of DFT and post-DFT methods. 
This has been discussed previously, and we can quote, for example, 
``Moreover, the physical parameters characterizing the electronic structure and interactions that constitute a key input into the theoretical descriptions of the unconventional magnetism have not yet been determined." and
``... determining the accurate values of physical parameters related to the Kitaev physics, possibly from electronic structure studies, is important." \cite{sinnElectronicStructureKitaev2016}.
    
\begin{table}[!htbp]
\centering
\caption{
Magnetic moments $\mu_z$  of Ru in $\alpha$-RuCl$_3$ from
the references and from this work.
}
\label{tab:moments}
\begin{tabular}{l|c}
\hline
Ref. & $\mu_z$ [$\mu_B$] \\
\hline
Banerjee et al. \cite{banerjeeProximateKitaevQuantum2016}
     & 0.4(1)   \\
Cao et al. \cite{caoLowtemperatureCrystalMagnetic2016}
     & 0.45(5) \\
Johnson et al. \cite{johnsonMonoclinicCrystalStructure2015}
     & 0.64(4)  \\
This work, PBE0(15\%)
     & 0.89     \\
This work, PBE0(20\%)
     & 0.91     \\
Ref. \cite{setyawan_high-throughput_2010, aflow}
     & 0.935 \\

\hline
\end{tabular}
\end{table}

\subsection{QMC Finite-Size Errors}
\label{sub:finite_size}
Finite-size errors can significantly affect total energies per particle and band gap estimates. 
In self-consistent approaches such as DFT or HF, these issues are easier to control in bulk solids using a very dense
$k-$mesh. In many-body methods such as QMC or even mean-field calculations with defects, one has to employ increasingly
large supercells to extrapolate to the thermodynamic limit (TDL). Often, finite-size corrections for kinetic
\cite{chiesaFiniteSizeErrorManyBody2006} and potential \cite{fraserFinitesizeEffectsCoulomb1996,
williamsonEliminationCoulombFinitesize1997, kentFinitesizeErrorsQuantum1999} energies are used in QMC since very large
supercells with fixed-node/fixed-phase DMC calculations can become prohibitively expensive.

\begin{figure}[htbp!]
    \centering
    \includegraphics[width=0.5\textwidth]{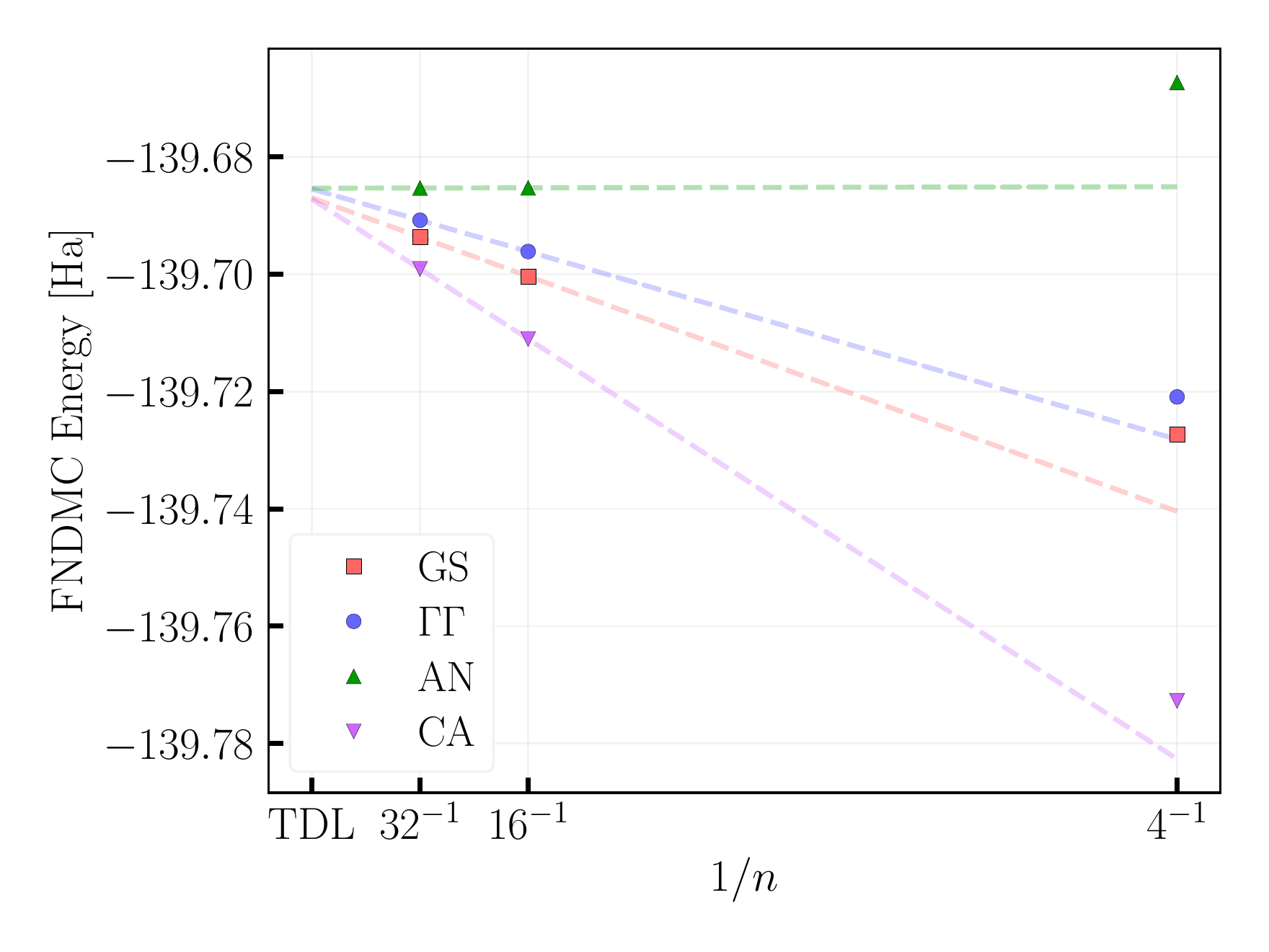}
    \caption{Extrapolations of DMC/PBE0(15\%) energy per chemical formula for $n=(16,32)$ in ground, $\Gamma\to\Gamma$
    excited, anionic, and cationic states.}
    \label{fig:fs_extrap}
\end{figure}

In this work, we opt for commonly used Ewald interactions for the potential energy and single $k-$point supercells, which
worked very well in previous calculations such as for Si \cite{annaberdiyevCohesionExcitationsDiamondstructure2021} and LaScO$_3$ \cite{meltonManybodyElectronicStructure2020} solids provided sufficiently large cells are used, and thermodynamic limit can be ascertained for all the states involved. In
\fref{fig:fs_extrap} we show the TDL extrapolations using DMC/PBE0(15\%) energies for all the states we calculated. The
data points correspond to the $[1\times1\times1]$ ($n=4$), $[2\times1\times2]$ ($n=16$), and  $[2\times2\times2]$ ($n=32$)
multiples of the conventional cell where $n$ is the number of RuCl$_3$ chemical formula units in a supercell. The $[2\times1\times2]$ ($n=16$) was particularly chosen for its near-cubic shape (see the Supplementary Material for visualizations of
the supercells). We use the largest two supercells for extrapolations, and the crucial point is the agreement of the
energy per formula for all the states within 2.4 mHa since one-electron excitation energies between the states must
vanish in the thermodynamic limit. From previous experience \cite{annaberdiyevCohesionExcitationsDiamondstructure2021,
meltonManybodyElectronicStructure2020} we did not include the smallest size in the extrapolations since we found that
this can be a source of additional bias instead of improvement. Too small supercells typically show shifts in both kinetic
and potential energies that do not conform to the asymptotic scaling and, therefore, can be counterproductive for
extrapolations. 
The observed agreement in \fref{fig:fs_extrap} gives us a clear indication that the supercells are large
enough for reliable estimations of desired quantities such as cohesive energy or energy gaps.

\subsection{Cohesive Energies}
\label{sub:cohesive}

\begin{figure}[htbp!]
    \centering
    \includegraphics[width=0.5\textwidth]{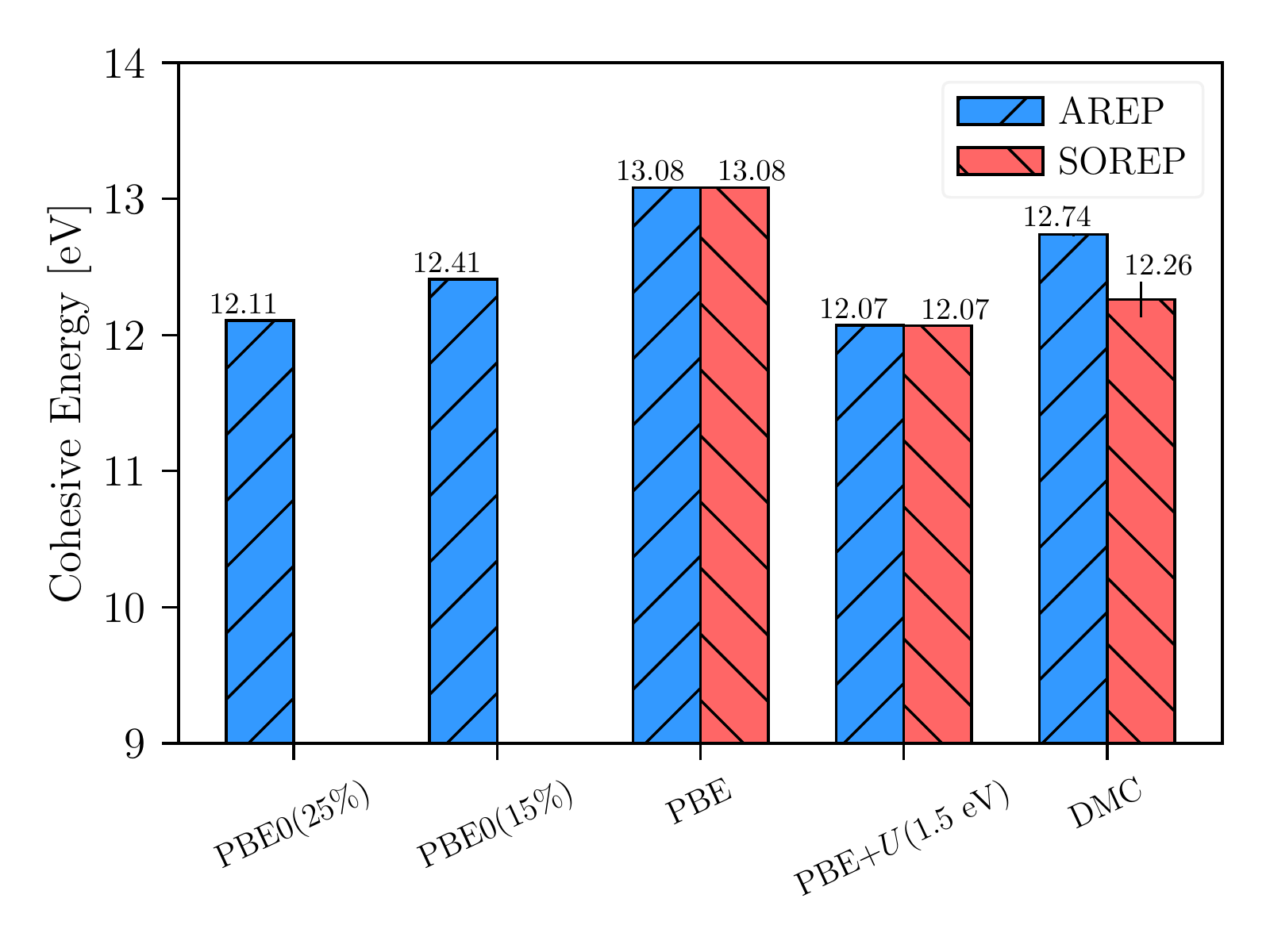}
    \caption{Cohesive energies using various methods using AREP and SOREP. 
    Above 9~eV is shown to better illustrate the differences in methods.
    }
    \label{fig:cohesive_energies}
\end{figure}

Let us start by focusing on the cohesive energy of $\alpha$-RuCl$_3$. \fref{fig:cohesive_energies} provides the cohesive
energies using various methods and estimations with AREP and SOREP. As described previously, the two largest supercells
$n=16$ and $n=32$ were used to estimate the QMC TDL cohesive energy. 
(The total energies for atoms, each supercell, and TDL can be found in the Supplemental Material). 
The AREP value for DMC uses the PBE0(15\%) as the reference, resulting in the lowest energy among all tested trial wave functions. 
Unfortunately, we were not able to properly converge the DFT/PBE0(15\%)
SOREP SCF calculations. Therefore, we used the DFT$+U$(1.5 eV) to converge the needed states in the SOREP setting and
subsequently, we estimate the changes due to spin-orbit coupling as follows:
\begin{multline}
    \label{eqn:cohesive}
    \rm \mathcal{E}^{SOREP}_{DMC/PBE0(15\%)} = \mathcal{E}^{AREP}_{DMC/PBE0(15\%)} + \\
    \rm + [\mathcal{E}^{SOREP}_{DMC/PBE+U(1.5eV)} - \mathcal{E}^{AREP}_{DMC/PBE+U(1.5eV)}]
\end{multline}
where $\mathcal{E}$ represents the cohesive energy, and the term in brackets represents cohesive energy change due to SOC.
We expect this estimator to introduce only very marginal bias since the energy differences using DMC/PBE0($\omega$) and
DMC/PBE$+U$ systematically agree as can be seen in \fref{fig:dmc_scan_avg_gap}. The biases are further diminished since
the SOC effect is evaluated as the difference of differences.

We probe two trial wave functions for atomic DMC energies: single-reference and proper multi-reference from the COSCI method.
If single-reference atomic energy is used, the cohesive energy is reduced by 0.48 eV, while the multi-reference case results in 0.61 eV reduction due to SOC.
We take the difference between the two references as a lower bound on the systematic error of the cohesive energy.
In \ref{fig:cohesive_energies}, we plot the DMC/SOREP using single-reference energies to achieve better error cancellation and the abovementioned systematic error of $\approx 0.1$ eV is shown.

The significant reduction of binding due to SOC has been observed before in several molecular cases for FNDMC/FPSODMC binding energies that are in excellent agreement with experiments \cite{melton_quantum_2016,melton_spin-orbit_2016}.
This follows from the fact that the energy gains from SOC in isolated atoms are larger than in solid state orbitals  that, at least in this material, appear to average out the SOC effects due to hybridization and periodicity.

To the best of our knowledge, an experimental cohesive energy estimate has not been published so far. It is worth noting
that our DMC estimate of cohesive energy [\textbf{12.3(1) eV}] closely agrees with DFT/PBE0(15\%) value and this also
provides another feedback for the optimal value of the exact-exchange mixing in the functional. For completeness, we
point out that DFT/PBE overestimates while DFT/PBE0 and DFT$+U$(1.5 eV) slightly underestimate the cohesive energies when
compared to DMC or DFT/PBE0(15\%) values, and this agrees with what has been seen in other systems \cite{shin_cohesion_2014, mcclainGaussianBasedCoupledClusterTheory2017, gruneisSecondorderMollerPlesset2010}.

\subsection{Band Gaps}
\label{sub:gaps}

Similar to the estimation of cohesive energy, for evaluation of band gaps, we rely on the largest two supercells we
were able to afford, with $n=16$ and $n=32$ units. 
One possible way to calculate $E_g$ and $E_G$ is to simply use the difference of extensive energies as in Eqns. \ref{eqn:optical_gap}, \ref{eqn:fundamental_gap}.
Unfortunately, this gets problematic due to the cost of required error bars and also due to the difficult estimation of systematic biases for total energies in large supercells. 
Clearly, using simple extensive energies for very large supercells becomes unreliable
since one tries to obtain a small gap from two large numbers with an increasing level of noise (in this case, the gap is
$\approx$ eV, whereas the largest supercell energy is $\approx -121640$ eV). Therefore, we decided to use another
approach that relies on energy per $n$ (intensive energies) and then uses the slopes of different states to evaluate
$E_g$ and $E_G$ in a manner that is consistent with the thermodynamic limit. Detailed discussions about using slopes to
obtain gaps and related elaborations can be found in Refs. \cite{annaberdiyevCohesionExcitationsDiamondstructure2021},
\cite{meltonManybodyElectronicStructure2020}.

\begin{table}[!htbp]
\centering
\caption{
DMC $\Gamma\to\Gamma$ excitonic ($E_g$) gaps [eV] using various trial wave functions.
}
\label{tab:gaps_Eg}
\begin{tabular}{l|ccc|ccc}
\hline
\hline
WF & $n=4$ & $n=16$ & $n=32$ & TDL \\
\hline
\multicolumn{5}{c}{AREP} \\
\hline
PBE0(15\%)      & 0.70(4) &  1.83(5) &  2.5(1) & 1.9(2) \\
PBE$+U$(1.5 eV) & 0.59(3) &  1.72(5) &  2.3(1) & 1.8(1) \\
\hline
\multicolumn{5}{c}{SOREP} \\
\hline
PBE$+U$(1.5 eV) & 0.97(4) &  1.98(7) &         &        \\
\hline
\hline
\end{tabular}
\end{table}

\begin{table}[!htbp]
\centering
\caption{
DMC $\Gamma\to\Gamma$ quasi-particle ($E_G$) gaps [eV] using various trial wave functions.
}
\label{tab:gaps_EG}
\begin{tabular}{l|ccc|ccc}
\hline
\hline
WF & $n=4$ & $n=16$ & $n=32$ & TDL \\
\hline
\multicolumn{5}{c}{AREP} \\
\hline
PBE0(15\%)      & 1.59(6) &   1.9(1) &  2.5(2) &  2.3(4) \\
PBE$+U$(1.5 eV) & 1.55(5) &  2.02(9) &  2.7(2) &  2.1(2) \\
\hline
\multicolumn{5}{c}{SOREP} \\
\hline
PBE$+U$(1.5 eV) & 1.64(7) &  2.2(1)  &         &         \\
\hline
\hline
\end{tabular}
\end{table}

Table \ref{tab:gaps_Eg} provides the excitonic gaps ($E_g$), and Table \ref{tab:gaps_EG} provides the fundamental gaps
($E_G$) calculated via DMC using hybrids and DFT$+U$ trial wave functions. We can draw several points from this
data. One is that $E_g \approx E_G$ within error bars is obtained considering the largest supercells as well as TDL
gaps. In fact, obtaining similar values for $E_g$ and $E_G$ is not surprising, and our $E_g$ values represent
fundamental gaps as discussed at length elsewhere, see Refs. \cite{dubeckyFundamentalGapFluorographene2020, perdew_understanding_2017}.
Note that, depending on the system, there are usually minor differences between these two estimators. This is caused by
a combination of two effects. One is the differences in systematic errors in charged vs. uncharged states since
these are not identical. Typically, the charged state convergence to the thermodynamic limit is slower due to the Ewald
model for potential energies.
The root cause here comes from the homogeneous charge compensation background that contrasts with
excitation-related charges, with added or
subtracted electronic states, which are inhomogeneous. Corrections of this model are not straightforward, especially in QMC, where very large supercells could be rather costly or even out of reach. Furthermore, the fixed-node errors for the considered states might not be uniform. There is a related bias, as can be seen, for example, in our calculations of excitations in simpler molecular systems
\cite{wang_binding_2020}.

Assuming $E_g \approx E_G$ and taking the average of TDL gap values from Tables \ref{tab:gaps_Eg} and \ref{tab:gaps_EG},
our estimate for AREP $E_G$ is \textbf{2.0(1) eV}. The SOC effect on gaps is extracted from the $n=16$ supercell
energies (the largest we were able to afford using SOREP). Once again, taking the average of Tables \ref{tab:gaps_Eg} and
\ref{tab:gaps_EG}, the SOC effect on gaps is an increase by \textbf{0.22(8) eV}. Therefore, our estimate for SOREP
$E_G$ is \textbf{2.2(1) eV} from DMC calculations. This is significantly larger than some previous band gap estimates
\cite{plumbEnsuremathAlphaEnsuremath2014, sandilandsSpinorbitExcitationsElectronic2016, kimKitaevMagnetismHoneycomb2015,
zhangTheoreticalStudyCrystal2021}, while it agrees with more recent results such as from Sinn et al
\cite{sinnElectronicStructureKitaev2016}. However, Sinn et al \cite{sinnElectronicStructureKitaev2016} estimate the
increase in the gap due to SOC to be $\approx 0.6$ eV while our estimate of $0.22(8)$ eV is much smaller.
We also note that PBE0(20\%), which is energetically close to the optimal PBE0(15\%) (Figure \ref{fig:dmc_scan_avg_tot}), provides a $\Gamma \to \Gamma$ gap value of $2.02$~eV which agrees with FNDMC value of 2.0(1)~eV in SOC-averaged framework.
On the other hand, we find that a larger value of $U \approx 4$~eV is required to reproduce the $\approx 2$~eV gap when compared to the optimal value of $U=1.5$~eV in the DFT$+U$ method.
For the sake of completeness, we point out that the gap is strongly overestimated in Hartree-Fock, with a value of $\approx11.7$~eV as seen in other  Mott-Hubbard antiferromagnets with transition elements. 

\begin{figure}[htbp!]
    \centering
    \includegraphics[width=0.5\textwidth]{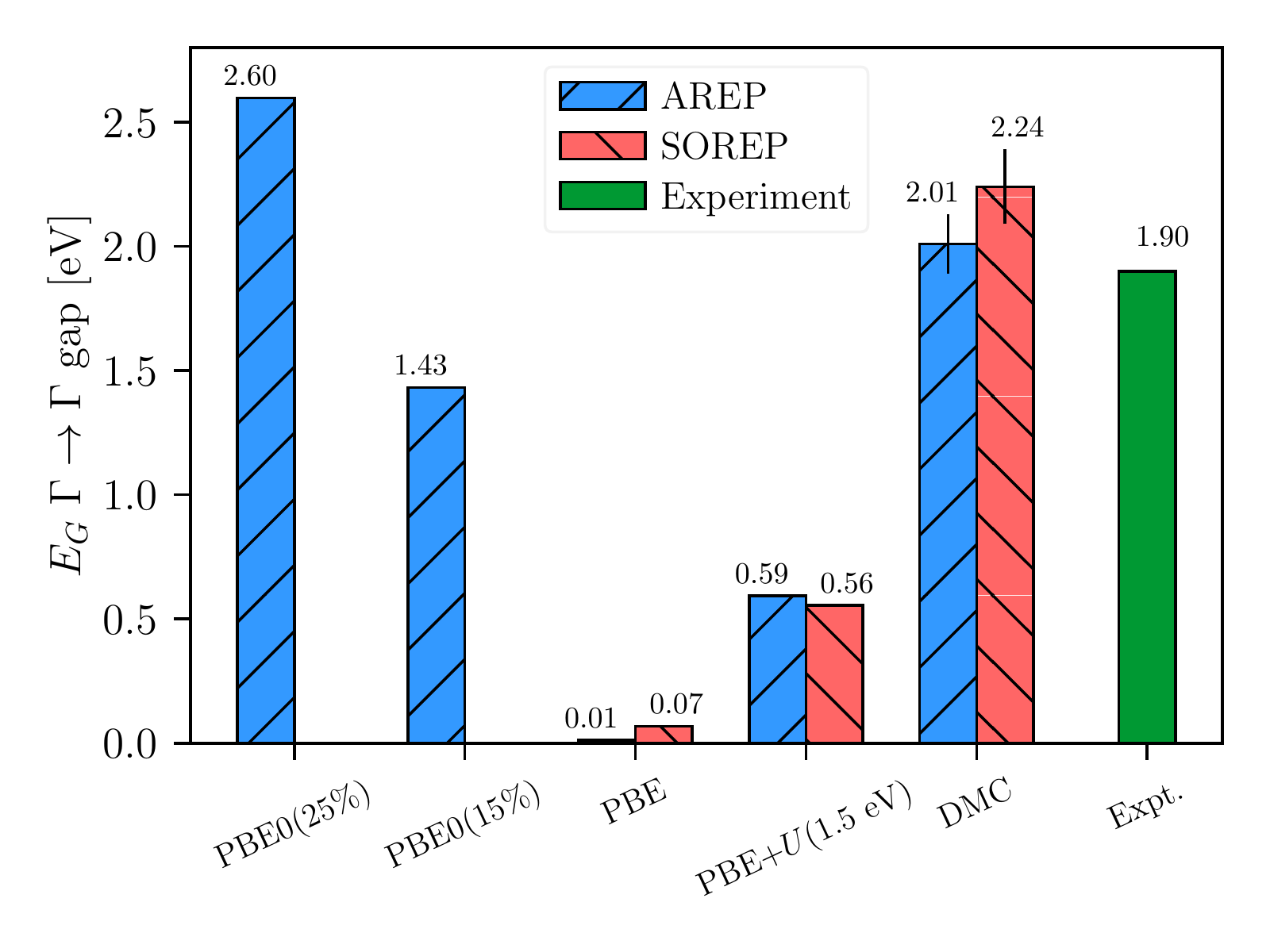}
    \caption{Gaps using various methods using AREP and SOREP.
    Experimental (Expt.) results are from Ref. \cite{sinnElectronicStructureKitaev2016, nevola_timescales_2021, Nevola}
    }
    \label{fig:gaps}
\end{figure}

Figure \ref{fig:gaps} summarizes the $E_G$ gaps obtained from the data we obtained in various methods. We observe the
usual underestimation of the band gap using PBE, while the hybrids exhibit values closer to DMC and to
experiment. Our DMC estimate $E_G=2.2(1)$ is in a good agreement with the experimental value of 1.9 eV obtained from
photoemission (PE), inverse-PE \cite{sinnElectronicStructureKitaev2016}, and angle-resolved photoemission (ARPES)
measurements \cite{nevola_timescales_2021, Nevola}. A small overestimation with respect to the experiment is typical for the
QMC data due to marginally larger fixed-node bias in excitations than in the ground states.  
Our results also indicate that the sharp $\alpha$ peak at around 1 eV observed in optical experiments \cite{plumbEnsuremathAlphaEnsuremath2014, sandilandsOpticalProbeHeisenbergKitaev2016, sandilandsSpinorbitExcitationsElectronic2016} has either an excitonic nature, or it is related 
to defects/impurities, or a combination thereof. Because $\alpha$-RuCl$_3$ is a quasi-2D material, this is not surprising since very large excitonic binding energies of that order are observed in other 2D materials \cite{zhengExcitonsTwoDimensionalMaterials2020, muellerExcitonPhysicsDevice2018}.
Indeed, recent time-resolved two-photon PE spectroscopy experiments indicate that abovementioned $\sim 1$ eV gap is excitonic \cite{nevola_timescales_2021}.

\section{Conclusions}
\label{sec:conclusions}
In this work, we presented highly accurate benchmark results for $\alpha$-RuCl$_3$ by studying its cohesive energy and
band gaps. In order to apply the many-body method to a complex material such as RuCl$_3$, we also advance several
methodological steps that lead to less ambiguous results and provide a more transparent path for
similar studies in the future. Except for the employed experimental geometry, this work provides a fully
\textit{ab-initio} study of $\alpha$-RuCl$_3$.

First, we use both spin-orbit averaged conventional fixed-node DMC method as well as two-component spinor-based
fixed-phase DMC, which explicitly includes the spin-orbit interaction. For this purpose, the atom of Cl is represented by
recently developed ccECPs with [Ne]-core. For Ru, we adopt ECP with valence $4s^24p^64d^65s^2$ from the Stuttgart table
\cite{petersonEnergyconsistentRelativisticPseudopotentials2007} with a minor adjustment that smooths out the effective
Coulomb term into a finite value at the ion origin. This straightforward modification keeps the accuracy of the original
construction essentially untouched. 

Second, we employ the spin-averaged framework both in DFT and in QMC methods which provides an initial picture of the
electronic structure, such as DFT bands, estimates of cohesive energies, and band gaps. We have made an effort to
estimate the extent of systematic errors such as the quality of the ECP, basis sets, finite-size effects, and
fixed-node/fixed-phase errors and to show relevant details of the calculations for maximum transparency.

Part of this is also a probe of the spin-orbit impact using both DFT with Hubbard $U$ and also two-component DMC
calculations.
Considering an ideal system without localized and strong excitonic effects, we do not see any significant changes in the cohesive energy and band gap estimates, and the obtained SOC-related shifts are mild,
in contrast with previous suggestions \cite{banerjeeProximateKitaevQuantum2016, eichstaedt_deriving_2019, sandilandsSpinorbitExcitationsElectronic2016, sears_magnetic_2015, sinnElectronicStructureKitaev2016, zhouAngleresolvedPhotoemissionStudy2016}. 
The SOC shifts of gaps appear to be bounded by $\approx$ 0.2 eV, in line with expectations for Ru from atomic and molecular calculations. Our estimates that
include both promotion (charge-neutral excitations) and fundamental (charged calculations with ionized supercells) gaps
show good agreement with PE/IPE and ARPES experiments with an overall accuracy of $0.2$ eV. To the best of our
knowledge, the experimental cohesive energy is not known, and our DMC result provides a truly many-body prediction for this
quantity.

We note that the much more subtle physics of the possible spin liquid phase remains hidden in the statistical noise. Clearly, for this purpose, one needs to build a proper effective Hamiltonian,
for the relevant low-lying states.
However, despite many attempts to do so previously in various settings, we are not much closer to a genuine understanding of the spin liquid aspects of this material. In particular, the spin liquid phase studies using DFT and downfolding approaches based on DFT-derived wave functions \cite{eichstaedt_deriving_2019, yamaji_first-principles_2014} require knowledge about suitable values of $U$ or exact-exchange mixing $\omega$ which are not known \textit{a priori}.
Such a study would also need to account for the role of phonons, as recent thermal Hall effect measurements showed that the thermal Hall signal is dominated by phonons \cite{lefrancois_evidence_2022, hentrich_unusual_2018}.
We believe that the insights provided in this work regarding the weak strength of SOC and the importance of electron correlations will be an excellent starting point for future studies in this direction. Our results show that hybrid DFT with $\omega$ exact-exchange mixing in the range of $15\%-20\%$ provides a representative, effective one-particle picture of
ideal $\alpha$-RuCl$_3$. Beyond this level, we conjecture that a more accurate correlation treatment with multi-reference wave functions will be needed to evaluate appropriate parameters for downfolding studies relevant to spin liquid phases.
Indeed, after solving the important challenge of multi-reference state construction and optimization in solids, this is the next opportunity for providing deeper insights based on spinor-based, many-body
wave functions.

From a broader perspective,
we believe the significance of the presented DMC results goes beyond only being relevant to this material.
In particular, the presented data reveal vital insights about appropriate DFT functionals for studies in systems with
heavier atoms and non-negligible spin-orbit effects. We observe, now in a more systematic manner \cite{bennettOriginMetalInsulatorTransitions2022}, that the nodes of trial
functions obtained from hybrid DFT are more accurate than for orbitals from DFT$+U$, suggesting thus that the hybrid
orbitals better describe the most optimal single-particle Hamiltonian.
Note that these two theories address the shortcomings of practically used DFT functionals
from two different sides. In hybrid functionals, the effects of local exchange are emphasized, which leads to lowering the
energy in the triplet vs. singlet channels. DMC calculations, being variational, indeed support this effect since that
leads to lower total energies by exchange-driven, better localization of orbitals. On the other hand, Hubbard $U$ pushes
up the singlet channel relative to the triplet channel. Although the relative effect of the singlet-triplet energy shift
is qualitatively similar, the hybrid functional electronic picture is closer to reality, at least for the purpose of providing a better variational orbital set for ground state calculations. Assuming that the excited states of interest are
essentially single-reference, similar conclusions would also apply to such excitation calculations. In case the
calculated state is not a single-reference, neither of these orbital sets would be optimal, and further processing would be needed, for example, by obtaining the natural orbitals from diagonalization of correlated single-particle density matrix \cite{obdmqmc1998}.

As far as we are aware, this is the first study of solids with spin-orbit effects using many-body QMC in explicit two-component spinor formalism.
Considering all the aspects of the presented work,  it provides important, many-body derived data for effective spin liquid studies and opens further opportunities for many-body wave function
studies of systems with heavy atoms and strong spin-orbit effects.

\section{Data Availability}
See Supplemental Material \cite{supplemental} for raw total energies, bulk geometry, and convergence
studies. The input, output files, and supporting data generated in this work are published in Materials Data
Facility~\cite{blaiszik_materials_2016, blaiszik_data_2019} and can be found in Ref.~\cite{mdf_data}.

The Department of Energy will provide public access to these results of federally sponsored research in accordance with the DOE Public Access Plan (http://energy.gov/downloads/doe-public-access-plan).

\begin{acknowledgments}

We would like to thank P. R. C. Kent for suggesting studying RuCl$_3$ material and for valuable comments. We also thank
Daniel Dougherty and Daniel Nevola for helpful and stimulating discussions. 
The authors are grateful to Joshua P. Townsend and Panchapakesan Ganesh for reading the manuscript and for helpful suggestions.

This work has been supported by the U.S. Department of Energy, Office of Science, Basic Energy Sciences, Materials
Sciences and Engineering Division, as part of the Computational Materials Sciences Program and Center for Predictive
Simulation of Functional Materials.
Part of this research (DFT calculations) was conducted by A. A. at the Center for Nanophase Materials Sciences (CNMS), which is a DOE Office of Science User Facility.

An award of computer time was provided by the Innovative and Novel Computational Impact on Theory and Experiment
(INCITE) program. This research used resources of the Oak Ridge Leadership Computing Facility, which is a DOE Office of
Science User Facility supported under Contract No. DE-AC05-00OR22725. This research used resources of the Argonne Leadership
Computing Facility, which is a DOE Office of Science User Facility supported under Contract No. DE-AC02-06CH11357. This
research used resources of the National Energy Research Scientific Computing Center (NERSC), a U.S. Department of Energy
Office of Science User Facility located at Lawrence Berkeley National Laboratory, operated under Contract No.
DE-AC02-05CH11231.

Sandia National Laboratories is a multimission laboratory managed and operated by National Technology \& Engineering
Solutions of Sandia, LLC, a wholly owned subsidiary of Honeywell International Inc., for the U.S. Department of Energy’s
National Nuclear Security Administration under Contract No. DE-NA0003525. Sandia National Laboratories is a multimission
laboratory managed and operated by National Technology \& Engineering Solutions of Sandia, LLC, a wholly owned
subsidiary of Honeywell International Inc., for the U.S. Department of Energy’s National Nuclear Security Administration
under Contract No. DE-NA0003525.

This paper describes objective technical results and analysis. Any subjective views or opinions that might be expressed
in the paper do not necessarily represent the views of the U.S. Department of Energy or the United States Government.

This manuscript has been authored by UT-Battelle, LLC under Contract No. DE-AC05-00OR22725 with the U.S. Department of Energy. The United States Government retains and the publisher, by accepting the article for publication, acknowledges that the United States Government retains a non-exclusive, paid-up, irrevocable, world-wide license to publish or reproduce the published form of this manuscript, or allow others to do so, for United States Government purposes.

\end{acknowledgments}



\UseRawInputEncoding
\bibliographystyle{apsrev4-2}
\bibliography{main}

\end{document}